\documentclass[conference]{IEEEtran}

\IEEEoverridecommandlockouts

\usepackage{cite}
\usepackage{amsmath,amssymb,amsfonts}
\usepackage{algorithmic}
\usepackage{graphicx}
\usepackage{subcaption}
\usepackage{textcomp}
\usepackage{xcolor}
\usepackage{comment}
\usepackage{bm}
\usepackage{mathtools}
\usepackage[hidelinks]{hyperref}

\usepackage{amsthm}
\newtheorem{theorem}{Theorem}

\newtheorem{proposition}{Proposition}

\newcommand{\euler}{\mathsf{e}}

\newcommand{\T}{^{\mathsf{T}}}
\newcommand{\R}[1]{\mathrm{#1}}

\newcommand{\B}[1]{\if#1\relax\bm{#1}\else\mathbf{#1}\fi} % bold text
\newcommand{\BB}[1]{\mathbb{#1}}
\newcommand{\C}[1]{\mathcal{#1}}

\renewcommand{\vec}[1]{\if#1\relax\bm{#1}\else\mathbf{#1}\fi} % vectors
\newcommand{\mat}[1]{\if#1\relax\bm{#1}\else\mathbf{#1}\fi}   % matrices

\def\BibTeX{{\rm B\kern-.05em{\sc i\kern-.025em b}\kern-.08em
    T\kern-.1667em\lower.7ex\hbox{E}\kern-.125emX}}

\begin{document}

\title{Survivability guarantees for delayed\\serial supply chains under demand shocks}

\author{
Lorenzo Giordano, Francesco Lo Iudice, Josep M. Olm, and Marco Coraggio%
\thanks{L. Giordano and M. Coraggio are with the Scuola Superiore Meridionale, Naples, Italy (e-mail: \texttt{l.giordano@ssmeridionale.it};
\texttt{m.coraggio@ssmeridionale.it}).}%
\thanks{F. Lo Iudice is with the Department of Electrical Engineering and Information Technologies, Università degli Studi di Napoli Federico II, Naples, Italy (e-mail: \texttt{francesco.loiudice2@unina.it}).}%
\thanks{J. M. Olm is with the Department of Mathematics and the Institute of Industrial and Control Engineering, Universitat Politècnica de Catalunya, Barcelona, Spain (e-mail: \texttt{josep.olm@upc.edu}). His work was partially supported by the Government of Spain through Project PID2025-172828OB-I00, and by the Generalitat de
Catalunya through Project 2025 SGR 00170.}%
}

\maketitle

\begin{abstract}
When demand exceeds the maximum production capacity of a supply chain, it can be satisfied only for a finite time, determined by the inventory stored along the chain.
Existing control methods ensure indefinite constraint satisfaction under bounded demand, but provide no guarantees on how long a supply chain can withstand a shock in which demand goes beyond production capacity.
We address this gap for a serial supply chain with transportation delays and flow constraints. Under a specified flow policy, we derive necessary and sufficient conditions on the inventory distribution that guarantee a prescribed lower bound on the exit time from the set where demand is served. 
We use these conditions to compute inventory targets that minimize holding costs via a quadratic program. 
Then, we propose a two-mode controller that builds reserves under ordinary demand and depletes them during the shock.
Simulations illustrate this approach.
\end{abstract}


\section{Introduction}
\label{sec:introduction}

Supply chains play a central role in the global economy but are increasingly exposed to disruptions, such as supply shortages or unexpected demand fluctuations. These events can propagate throughout the network, causing increased costs, delivery delays, and service interruptions. Recent crises, such as the COVID-19 pandemic \cite{ivanov_2020b}, have highlighted the vulnerability of supply chains to such disruptions, emphasizing the need for control strategies that improve their robustness.

Various definitions of robustness exist in the supply chain management literature. Specifically, in \cite{durach_2015}, it was characterized as the ability of the supply chain to resist or avoid changes caused by disturbances. Meanwhile, in \cite{monostori_2018} and \cite{monostori_2021}, the emphasis was on performance acceptability, describing a robust supply chain as one whose relevant key performance indicators remain within a predefined robustness zone during and after disruptions.

From a control-theoretic perspective, robustness can be formulated in terms of robust positive invariance of a set defined by operational constraints. Indeed, analytical conditions and feedback control strategies for satisfying an uncertain but bounded demand while fulfilling inventory constraints were derived for both discrete-time \cite{blanchini_2000, borrelli_2009} and continuous-time \cite{farraa_2021} models of production--distribution systems. Moreover, MPC strategies have been proposed for supply chains \cite{alessandri_2011} and sparse storage networks \cite{lejarza_2021}.
However, these approaches primarily aim at indefinite constraint satisfaction, which may become impossible when demand exceeds the maximum supply capacity for long enough time periods, eventually leading to inventory depletion. In such cases, the key questions are how long the supply chain can satisfy the demand using available inventory and how to prepare to withstand a demand shock of a prescribed duration.

This finite-horizon perspective is related to the time-to-survive concept, introduced in supply chain risk management to quantify how long a supply chain can continue satisfying demand following a node failure \cite{simchi-levi_2015}. In control theory, related questions have been investigated through exit time and drift-counteraction optimal control problems, where the objective is to maximize the time before constraint violation \cite{kolmanovsky_2008, zidek_2017}. Nevertheless, these approaches do not provide conditions guaranteeing the satisfaction of operational constraints over a prescribed time horizon for continuous-time models with transportation delays. Motivated by this gap, we investigate finite-horizon demand satisfaction in delayed serial supply chains subject to demand shocks exceeding the maximum input capacity. 

The main contributions of this paper are twofold. First, by exploiting the structure of the delayed supply chain, we derive an analytical characterization of the inventory trajectories under a specified flow policy and establish necessary and sufficient conditions on the inventories for guaranteeing a prescribed survival time, thus providing a lower bound on the exit time from the operational set. Second, we use these conditions to find an inventory distribution that minimizes holding costs, and we develop a two-mode control strategy that regulates inventories during nominal operation and exploits the accumulated reserves during demand shocks. Finally, numerical simulations illustrate the proposed approach.

%---------------------------------
\section{Modeling}
\label{sec:modeling}

\begin{figure}
    \centering
    \includegraphics[width=0.75\linewidth]{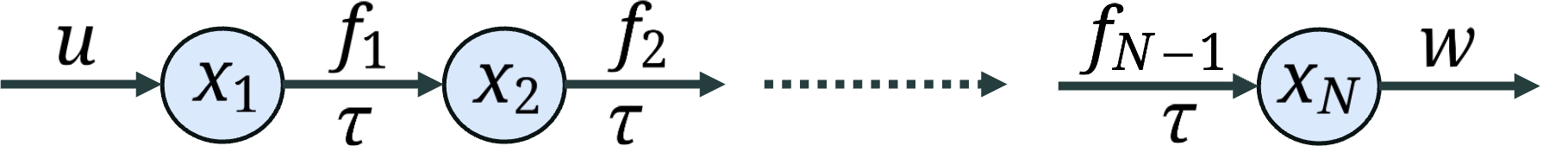}
    \caption{Schematic of a serial supply chain with $N$ nodes, external input $u$, internal flows $f_i$, transportation delays $\tau$, and external outflow $w$.}
    \label{fig:linear_graph}
\end{figure}

We consider a supply chain model where a single commodity is transferred sequentially and unidirectionally through a series of interconnected facilities, such as production centers, warehouses, and retailers, with transportation delays, similarly to \cite{farraa_2021}.
Specifically, we assume a directed linear graph with $N \ge 2$ nodes, one source (the production facility) and one sink (the final distributor), as illustrated in Fig.~\ref{fig:linear_graph}.
The chain dynamics is described by
\begin{subequations} \label{eq:chain_dynamics}
\begin{align}
\dot{x}_1(t) &= u(t)-f_1(t), \label{eq:chain_dynamics-a} \\
\dot{x}_i(t) &= f_{i-1}(t-\tau)-f_i(t), \quad i=2,\ldots,N-1, \label{eq:chain_dynamics-b} \\
\dot{x}_N(t) &= f_{N-1}(t-\tau) - w(t), \label{eq:chain_dynamics-c}
\end{align}
\end{subequations}
\begin{equation} \label{eq:outflow_rate}
    w(t) = \begin{cases}
    d(t), & \text{if } x_N(t)>0, \\
    0, & \text{if } x_N(t)=0,
    \end{cases}
\end{equation}
where $x_i(t)$ is the inventory at node $i$ and time $t$, 
$u(t) \in [0, u_{\R{max}}]$ is the controlled external inflow rate, 
$f_{i}(t)$ is the controlled commodity flow leaving node $i$ towards node $i+1$ at time $t$,
$d(t) \in [d_{\R{min}}, d_{\R{max}}]$ is the external demand rate,
$w(t)$ is the external outflow rate, and $\tau \in \BB{R}_{>0}$ is a transportation delay.

Let us define $\B{x} \coloneqq [ x_1 \ \cdots \ x_N ]\T$ and $\B{f} \coloneqq [ f_1 \ \cdots \ f_{N-1} ]\T$. 
Since~\eqref{eq:chain_dynamics} has delayed inputs, its evolution for $t \ge t_0$ is determined not only by $\B{x}(t_0)\in\BB{R}^N_{\ge 0}$, but also by the flow histories $\B{\eta}_{t_0}(\theta) \coloneqq \B{f}(t_0 + \theta)$ for $\theta\in[-\tau,0)$.
Moreover, \eqref{eq:outflow_rate} reflects that demand can be served only when inventory at the last node is positive.%
\footnote{The outflow law \eqref{eq:outflow_rate} is discontinuous at $x_N=0$.
This makes system \eqref{eq:chain_dynamics}--\eqref{eq:outflow_rate} \emph{piecewise-smooth}, implying solutions in the classical sense may not admit a continuation beyond this boundary; in that case, the Filippov notion of solution can be used.
However, our analysis concerns only trajectories with $x_N>0$ up to their first boundary contact, which is well defined by continuity of the state.}%

We assume that
(i) the function $d$ is piecewise constant, unknown in advance, but measurable in real-time,
(ii) $0\leq d_{\R{min}}<u_{\R{max}}<d_{\R{max}}$, and
(iii) $0 \le f_i(t) \le \beta x_i(t)$, 
where $\beta \in \mathbb{R}_{>0}$ is the maximum release rate, modeling a stock-dependent push release mechanism \cite{taboada_2022}. Assumption (iii), together with \eqref{eq:outflow_rate}, $\vec{x}(t_0) \ge 0$, and $\vec{\eta}_{t_0} \ge 0$, ensures non-negativity of the state, i.e., $\vec{x}(t) \ge 0$ for all $t\ge t_0$.

We define the \emph{operational set} $\mathcal{X}_{\mathrm{op}} = \{ \mathbf{x} \in \mathbb{R}_{\ge 0}^N : x_N > 0 \}$.
According to \eqref{eq:outflow_rate}, demand is fully served at time $t$ if $\vec{x}(t) \in \mathcal{X}_{\mathrm{op}}$.
For a given solution with $\vec{x}(t_0) \in \C{X}_{\R{op}}$, we define the \emph{exit time} (from the set $\C{X}_{\R{op}}$) as $T_{\R{exit}} \coloneqq \inf\{t\ge t_0 \mid \vec{x}(t) \notin \C{X}_{\R{op}}\} - t_0$ (with the convention that $\inf \varnothing = \infty$).

We refer to \emph{ordinary demand} when $d(t)\leq u_{\R{max}}$ and to a \emph{demand shock} when $d(t)> u_{\R{max}}$; in particular, we refer to \emph{constant} demand shocks of duration $T_{\R{s}}$ and magnitude $\bar{d} > u_{\R{max}}$ as conditions in which $d(t) = \bar{d}$ for all $t \in [t_0, t_0 + T_{\R{s}}]$, where $t_0$ is the onset time of the shock.
It is immediate to verify%
\footnote{The total amount of commodity in the chain $M_0(t)\coloneq\sum_{i=1}^{N}x_i(t)+\sum_{i=1}^{N-1}\int_{t-\tau}^{t}f_i(s)\,\R{d}s$ (which is non-negative since $x_i,f_i\ge 0$) satisfies $\dot M=u-w$.
Let $d(t)\ge\bar{d}>u_{\R{max}}$ for all $t\ge t_0$, and assume by contradiction that $\B{x}(t)\in\C{X}_{\R{op}}$ for all $t\ge t_0$ (i.e., $T_{\R{exit}}=\infty$).
Then $w=d$, so $\dot{M}_0 \le u_{\R{max}}-\bar{d}<0$ for all $t\ge t_0$, and $M_0$ eventually becomes negative in finite time, yielding a contradiction.} 
that if a demand shock persists indefinitely, no admissible $u$ and $\vec{f}$ can keep $\vec{x}(t) \in \C{X}_{\R{op}}$ for all $t \ge t_0$, i.e., $T_{\R{exit}} < \infty$. Then, we say that the chain (starting with $\B{x}(t_0)$ and $\B{\eta}_{t_0}$) \emph{survives} a demand shock of duration $T_{\R{s}}$ and magnitude $\bar{d}$ if, under a given control policy, $T_{\R{exit}} > T_{\R{s}}$.

Our goals are as follows: (i) determine reference inventory values $\vec{x}^\star \in \BB{R}^N$ that allow the chain to survive demand shocks of prescribed duration and magnitude, and (ii) design a control strategy that achieves $\vec{x}^\star$ under ordinary constant demand.

%------------------------------------
\section{Exit time characterization}
\label{sec:exit_time_characterization}

In this section, we study how the distribution of inventory across the supply chain affects its ability to withstand a demand shock. First, in Section~\ref{subsec:dynamics_with_maximum_flows}, we derive an analytical expression for the evolution of the inventories under a maximum-flow policy starting from a constant history. Then, in Section~\ref{subsec:lower_bound_exit_time}, we establish necessary and sufficient conditions on the inventory levels that guarantee a prescribed lower bound on the exit time under the same assumptions.

%-------------------------------
\subsection{Dynamics of the chain with maximum flow rates}
\label{subsec:dynamics_with_maximum_flows}

During a demand shock, the external input $u$ is no longer sufficient to meet the demand $d$, and the chain must rely on its stored inventories $\vec{x}$. However, inventory stored at upstream nodes cannot immediately contribute to demand satisfaction due to transportation delays $\tau$. Therefore, to facilitate the transfer of these reserves toward the final distributor, we consider a flow policy in which each node releases inventory at its maximum admissible rate, namely $f_i(t)=\beta x_i(t)$, for $i=1,\ldots,N-1$. Under this policy, and as long as the chain remains in $\mathcal{X}_{\mathrm{op}}$, model~\eqref{eq:chain_dynamics} reduces to
\begin{subequations} \label{eq:chain_dynamics_push}
\begin{align}
\dot x_1(t) &= u(t) -\beta x_1(t), \\
\dot x_i(t) &= \beta x_{i-1}(t-\tau) - \beta x_i(t), \quad i = 2, \dots, N-1, \\
\dot x_N(t) &= \beta x_{N-1}(t-\tau) - d(t).
\end{align}
\end{subequations}
To obtain a compact representation of the trajectories of \eqref{eq:chain_dynamics_push}, we define the functions $\phi_m, \psi_n : \BB{R} \to \BB{R}$, with $n, m \in \BB{N}_{\ge 0}$, given by
\begin{equation}\label{eq:erlang_kernel}
  \phi_m(\theta) \coloneq
  \begin{dcases}
    0, &\theta < 0,\\
    \frac{\beta^m}{m!}\, \theta^m\, \euler^{-\beta \theta}, &\theta \ge 0,
  \end{dcases}
\end{equation}
\begin{equation}\label{eq:erlang_cdf}
  \psi_{n}(\theta) \coloneq
  \begin{dcases}
    0, &\theta < 0,\\
    1 - \sum_{m=0}^{n-1} \phi_m(\theta), &\theta \ge 0.
  \end{dcases}
\end{equation}
In particular, for $\theta \ge 0$, $\phi_0(\theta) = \euler^{-\beta \theta}$ and $\psi_0(\theta) = 1$, with the empty sum understood to be zero. 
Note that, for $n \ge 1$, the functions $\beta\phi_{n-1}$ and $\psi_n$ correspond, respectively, to the probability density function and cumulative distribution function of an Erlang distribution with shape parameter $n$ and rate parameter $\beta$.
The corresponding Laplace transforms are
\begin{equation}\label{eq:transforms}
  \mathcal{L}\{\phi_m\}(s) = \frac{\beta^m}{(s+\beta)^{m+1}}, \quad
  \mathcal{L}\{\psi_n\}(s) = \frac{1}{s}\left(\frac{\beta}{s+\beta}\right)^{\!n}.
\end{equation}
The following proposition characterizes the response of the chain with the maximum-flow policy, i.e. \eqref{eq:chain_dynamics_push}, assuming constant flow histories $\vec{\eta}_0$ (cf.~Sec.~\ref{sec:modeling}).

\begin{proposition}\label{prop:response}
    Consider system \eqref{eq:chain_dynamics_push} with $\vec{x}(0)=\vec{x}_0 = [x_{0,1} \ \cdots \ x_{0,N}]\T$ and $\vec{\eta}_0 = \bar{\eta}\vec{1}_{N-1}$, $\bar{\eta} \in \BB{R}_{\ge 0}$.
    Assume that the input $u$ and demand $d$ are locally integrable and of exponential order. 
    Then, the inventory trajectories for $t \geq 0$ are:
    \begin{equation} \label{eq:general_response_internal}
    \begin{aligned}
    x_i(t) = {}&\frac{\bar{\eta}}{\beta} + \frac{1}{\beta}\sum_{k=1}^{i}(\beta x_{0,k}-\bar{\eta}) \phi_{i-k}\left(t-(i-k)\tau\right) \\
    &+\int_0^t \phi_{i-1}\left(t-(i-1)\tau-\sigma\right) \left(u(\sigma)-\bar{\eta}\right) \,\mathrm{d}\sigma,
    \end{aligned}
    \end{equation}
    for $i=1,\ldots,N-1$, and
    \begin{equation} \label{eq:general_response_terminal}
    \begin{aligned}
    x_N(t)={}&x_{0,N} +\frac{1}{\beta}\sum_{k=1}^{N-1}(\beta x_{0,k}-\bar{\eta}) \psi_{N-k}\left(t-(N-k)\tau\right) \\
    &+\int_0^t \psi_{N-1}\left(t-(N-1)\tau-\sigma\right)\left(u(\sigma)-\bar{\eta}\right)\,\mathrm{d}\sigma \\
    &-\int_0^t \left(d(\sigma)-\bar{\eta}\right) \,\mathrm{d}\sigma.
    \end{aligned}
    \end{equation}
\end{proposition}

\begin{proof}
Define the change of variables 
\begin{subequations}
\begin{align}
    v(t) &\coloneq u(t)-\bar{\eta}, \\
    z_i(t) &\coloneq \beta x_i(t)-\bar{\eta}, \quad i=1,\ldots,N-1, \label{eq:z_def}
\end{align}
\end{subequations}
The dynamics of \eqref{eq:z_def} are given by
\begin{align*}
\dot z_1(t) &= \beta v(t) - \beta z_1(t), \\
\dot z_i(t) &= \beta z_{i-1}(t-\tau) -\beta z_i(t), \quad i=2,\ldots,N-1.
\end{align*}
For $t \in [-\tau,0)$, the variables \eqref{eq:z_def} are defined from the prescribed flow histories, as $z_i(t)=f_i(t)-\bar{\eta}=0$, for all $t \in [-\tau,0)$.
Taking Laplace transforms yields
\begin{align*}
(s+\beta)Z_1(s) &=z_1(0)+ \beta V(s), \\
(s+\beta)Z_i(s) &=z_i(0)+\beta e^{-s\tau}Z_{i-1}(s),
\end{align*}
for $i = 2,\ldots, N-1$, 
where $Z_i$ and $V$ denote the Laplace transforms of $z_i$ and $v$.
Solving recursively results in
\begin{equation} \label{eq:Z_i_laplace}
Z_i(s) = \sum_{k=1}^{i} \frac{\beta^{i-k}e^{-(i-k)s\tau}}{(s+\beta)^{i-k+1}}z_k(0) + \frac{\beta^{i} e^{-(i-1)s\tau}}{(s+\beta)^i}V(s).
\end{equation}
Then, using \eqref{eq:transforms}, the time-shift property of the Laplace transform, and \eqref{eq:z_def} yields \eqref{eq:general_response_internal}.
As per the last inventory, its dynamics is
\begin{equation*}
    \dot x_N(t) = z_{N-1}(t-\tau) + \bar{\eta} - d(t).
\end{equation*}
Taking Laplace transforms yields
\begin{equation*}
    sX_N(s)-x_{0,N}= e^{-s\tau}Z_{N-1}(s) - \left(D(s)-\frac{\bar{\eta}}{s}\right),
\end{equation*}
where $X_N$ and $D$ denote the Laplace transforms of $x_N$ and $d$.
By substituting the expression for $Z_{N-1}(s)$ from \eqref{eq:Z_i_laplace} and using \eqref{eq:transforms}, the inverse Laplace transform gives \eqref{eq:general_response_terminal}.
\end{proof}

Equation~\eqref{eq:general_response_terminal} highlights the different contributions of the initial inventories $\vec{x}_0$, external input $u$, and demand $d$ to the final inventory $x_N$.
In particular, the effect of each upstream inventory is determined by its position along the chain and the corresponding transportation delays. Similarly, variations in the external input affect the final inventory only after the total transportation delay $(N-1)\tau$, whereas demand variations have an immediate effect. Therefore, the response of the final inventory depends not only on the total amount of stored commodity, but also on its distribution across the chain.

%----------------------------------
\subsection{Lower bound on exit time}
\label{subsec:lower_bound_exit_time}

Next, in the case of a constant demand shock, assumed without loss of generality to occur at $t_0 = 0$, we derive necessary and sufficient conditions on the pre-shock inventory distribution that guarantee demand satisfaction for a prescribed horizon. 

\begin{theorem}
\label{thm:inventory_conditions}
Consider system \eqref{eq:chain_dynamics_push} with $\vec{x}(0)=\mathbf{x}_0$ and $\vec{\eta}_0 = \bar{\eta}\vec{1}_{N-1}$, $\bar{\eta} \in [0,u_{\R{max}}]$.
For $t > 0$, let $u(t) = \bar{u} \in [0, u_{\max}]$ and $d(t) = \bar{d} \in (u_{\max}, d_{\max}]$. 
Define $\Delta u \coloneq \bar{u} - \bar{\eta}$ and $\Delta d \coloneq \bar{d} - \bar{\eta}$.
Then, for any $T>0$, the exit time satisfies
\begin{equation}
    T_{\R{exit}} > \bar{T} = (N-1)\tau + T,
\end{equation}
if and only if the $N$ following inequalities hold:
\begin{equation}
\label{eq:inventory_N}
x_{0,N} > \Delta d \, \tau,
\end{equation}
\begin{equation} \label{eq:inventory_i}
\begin{aligned}
x_{0,i}>{}&\frac{\bar{\eta}}{\beta} +\sup_{\theta\in(0,\tau]} \frac{1}{\psi_{N-i}(\theta)} \Bigg[\Delta d\big(\theta+(N- i)\tau\big)-x_{0,N} \\
&-\frac{1}{\beta} \sum_{k=i+1}^{N-1} (\beta x_{0,k}-\bar{\eta}) \psi_{N-k}\left(\theta+(k-i)\tau\right) \Bigg],
\end{aligned}
\end{equation}
for $i=N-1,\ldots,2$, and
\begin{equation} \label{eq:inventory_1}
\begin{aligned}
x_{0,1}>{}&\frac{\bar{\eta}}{\beta} +\sup_{\theta\in(0,T]} \frac{1}{\psi_{N-1}(\theta)} \Bigg[ \Delta d(N-1)\tau \\
&+\frac{\Delta u}{\beta} \sum_{m=1}^{N-1}\psi_m(\theta) +(\Delta d-\Delta u)\theta-x_{0,N} \\
&-\frac{1}{\beta} \sum_{k=2}^{N-1} (\beta x_{0,k}-\bar{\eta}) \psi_{N-k}\left(\theta+(k-1)\tau\right) \Bigg].
\end{aligned}
\end{equation}
\end{theorem}

\begin{proof}
Particularizing \eqref{eq:general_response_terminal} to the specified post-shock input $\bar{u}$ and demand $\bar{d}$ yields
\begin{equation} \label{eq:xN_response}
\begin{aligned}
x_N(t) ={}& x_{0,N}
+\frac{1}{\beta}\sum_{k=1}^{N-1}
(\beta x_{0,k}-\bar{\eta})
\psi_{N-k}\left(t-(N-k)\tau\right)\\
&+\Delta u\int_0^t
\psi_{N-1}\left(t-(N-1)\tau-\sigma\right)
\,\mathrm{d}\sigma
-\Delta d\,t.
\end{aligned}
\end{equation}
First, consider $t\in[0,(N-1)\tau]$. Since $\psi_{N-1}(\theta)=0$ for $\theta<0$, the integral term in \eqref{eq:xN_response} vanishes throughout this interval. Partitioning it into $((N-i)\tau,(N-i+1)\tau]$, $i=N,\ldots,2$, and setting $t=(N-i)\tau+\theta$, with $\theta\in(0,\tau]$, gives
\begin{equation*} \label{eq:x_N_traslated_no_input}
\begin{aligned}
x_N(t) ={}& x_{0,N} +\frac{1}{\beta}\sum_{k=i}^{N-1} (\beta x_{0,k}-\bar{\eta}) \psi_{N-k}\left(\theta+(k-i)\tau\right) \\
&-\Delta d\big(\theta+(N-i)\tau\big).
\end{aligned}
\end{equation*}
For $i=N$, the sum is empty, and
\[
x_N(t)=x_{0,N}-\Delta d\,t, \quad t\in[0,\tau].
\]
Since $\Delta d>0$, positivity of $x_N(t)$ over this interval is equivalent to \eqref{eq:inventory_N}.

For $i=N-1,\ldots,2$, imposing $x_N(t)>0$ and separating the contribution of $x_{0,i}$ yields
\begin{equation*}
\begin{aligned}
x_{0,i} >{}& \frac{\bar{\eta}}{\beta} +\frac{1}{\psi_{N-i}(\theta)} \Bigg[ \Delta d\big(\theta+(N-i)\tau\big)-x_{0,N} \\
&-\frac{1}{\beta} \sum_{k=i+1}^{N-1} (\beta x_{0,k}-\bar{\eta}) \psi_{N-k}\left(\theta+(k-i)\tau\right) \Bigg],
\end{aligned}
\end{equation*}
for every $\theta\in(0,\tau]$. Since $\psi_{N-i}(\theta)>0$ for $\theta>0$, the division is well defined. By considering the supremum of the right-hand side over $(0,\tau]$, this expression reduces to \eqref{eq:inventory_i}.

Next, consider $t\in((N-1)\tau,(N-1)\tau+T]$. Setting $t=(N-1)\tau+\theta$, with $\theta\in(0,T]$, and using $\int_0^\theta\psi_{N-1}(s)\,\mathrm{d}s = \theta-\frac{1}{\beta}\sum_{m=1}^{N-1}\psi_m(\theta)$, equation~\eqref{eq:xN_response} becomes
\begin{equation*} \label{eq:x_N_traslated_with_input}
\begin{aligned}
x_N(t) ={}& x_{0,N} +\frac{1}{\beta}\sum_{k=1}^{N-1} (\beta x_{0,k}-\bar{\eta}) \psi_{N-k}\left(\theta+(k-1)\tau\right) \\
&+\Delta u\,\theta -\frac{\Delta u}{\beta} \sum_{m=1}^{N-1}\psi_m(\theta) -\Delta d\big((N-1)\tau+\theta\big).
\end{aligned}
\end{equation*}
Imposing $x_N(t)>0$ and separating the contribution of $x_{0,1}$ yields
\begin{equation*}
\begin{aligned}
x_{0,1} >{}& \frac{\bar{\eta}}{\beta} +\frac{1}{\psi_{N-1}(\theta)} \Bigg[ \Delta d(N-1)\tau \\
&+\frac{\Delta u}{\beta} \sum_{m=1}^{N-1}\psi_m(\theta) +(\Delta d-\Delta u)\theta-x_{0,N} \\
&-\frac{1}{\beta} \sum_{k=2}^{N-1} (\beta x_{0,k}-\bar{\eta}) \psi_{N-k}\left(\theta+(k-1)\tau\right) \Bigg],
\end{aligned}
\end{equation*}
for every $\theta\in(0,T]$. Since $\psi_{N-1}(\theta)>0$ for $\theta>0$, the division is well defined. Then, by taking the supremum of the right-hand side over $(0,T]$, \eqref{eq:inventory_1} follows.

Combining the conditions over all intervals establishes their necessity and sufficiency for $x_N(t)>0$, $t \in [0,\bar{T}]$, thus implying that $T_{\R{exit}} > \bar{T}$.
\end{proof}

Theorem~\ref{thm:inventory_conditions} provides a recursive characterization of the initial inventory levels required to sustain a constant demand shock $\bar{d}$ over a prescribed time horizon, assuming constant flow histories $\bar{\eta}$. 
Due to transportation delays, the external input cannot immediately replenish the final inventory; therefore, conditions $\eqref{eq:inventory_N}$ and $\eqref{eq:inventory_i}$ determine the inventory requirements before the external input reaches the final node, whereas condition $\eqref{eq:inventory_1}$ also accounts for its contribution.

 %------------------------------
\section{Control design}
\label{sec:control_design}

In this section, we develop a control strategy to prepare the supply chain for potential demand shocks.
First, in Section~\ref{subsec:optimal_selection}, we formulate an optimization problem to determine target inventory levels that ensure at least a prescribed exit time while minimizing holding costs. 
Then, in Section~\ref{subsec:control_strategy}, we design the controller to achieve these targets during nominal operation and sustain demand during a shock.

%------------------------------------------
\subsection{Optimal inventory level selection}
\label{subsec:optimal_selection}

Since in general infinitely many inventory distributions $\vec{x}_0$ satisfy the conditions of Theorem~\ref{thm:inventory_conditions}, we seek one that minimizes an inventory holding cost. In particular, we consider the worst-case demand increase from $d_{\R{min}}$ to $d_{\R{max}}$, assuming constant initial flow histories equal to $d_{\R{min}}$ and setting the external input to $u_{\R{max}}$ during the shock.

The conditions of Theorem~\ref{thm:inventory_conditions} provide the constraints for the optimization problem. While condition \eqref{eq:inventory_N} can be imposed directly, conditions \eqref{eq:inventory_i} and \eqref{eq:inventory_1} involve suprema over continuous intervals. To obtain a tractable formulation, we approximate them by enforcing the corresponding inequalities at finitely many sampling instants.
Specifically, for $i=1,\ldots,N-1$, we express conditions \eqref{eq:inventory_i} and \eqref{eq:inventory_1} as pointwise inequalities over the intervals $I_1=(0,T]$ and $I_i=(0,\tau]$ for $i=2,\ldots,N-1$. Then, we define $g_i(\theta,\mathbf{x}_0)$ by moving all terms of the corresponding inequality to the left-hand side, so that the conditions become $g_i(\theta,\mathbf{x}_0)>0$ for all $\theta\in I_i$.
Finally, we introduce finite sampling grids $\mathcal{T}_i=\{\theta_i^1,\ldots,\theta_i^{L_i}\}\subset I_i$ and impose $g_i(\theta_i^j,\mathbf{x}_0)\geq\varepsilon$, for all $j = 1, \dots, L_i$, where $\varepsilon>0$ is a prescribed safety margin.
Moreover, to ensure that each intermediate node can sustain a release rate up to $u_{\R{max}}$ when operating at its target inventory level, we impose $\beta x_{0,i}\geq u_{\R{max}}$, for $i=1,\ldots,N-1$.

The target inventory distribution is then obtained by solving
\begin{equation}
\label{eq:optimization}
\begin{aligned}
\mathbf{x}^\star = \arg\min_{\mathbf{x}_0\in\mathbb{R}_{\geq0}^N} 
\quad &\sum_{i=1}^N h_i x_{0,i}^2,\\
\text{s.t.}\quad
&g_i(\theta_i^j,\mathbf{x}_0)\geq\varepsilon,
&&\substack{i=1,\ldots,N-1, \\ j=1,\ldots,L_i,}\\
&x_{0,N}\geq\Delta d\,\tau+\varepsilon,\\
&\beta x_{0,i}\geq u_{\R{max}},
&&i=1,\ldots,N-1,
\end{aligned}
\end{equation}
where $\Delta d=d_{\R{max}}-d_{\R{min}}$ and $h_i>0$, $i=1,\ldots,N$, are the inventory holding cost coefficients. Since the functions $g_i(\theta, \mathbf{x}_0)$ are affine in $\mathbf{x}_0$ for each fixed $\theta_i^j$, the resulting problem is a convex quadratic program. Its solution defines the target inventory distribution to be established during nominal operation in preparation for a possible demand shock.

The discretization provides a finite-dimensional approximation of the conditions in Theorem~\ref{thm:inventory_conditions}. However, the approximation can be refined by increasing the number of sampling points or using constraint generation techniques.

\subsection{Control strategy}
\label{subsec:control_strategy}

We now design a control strategy for the supply chain \eqref{eq:chain_dynamics} that determines the external input $u(t)$ and the internal flows $f_i(t)$ with two objectives: when an ordinary demand is observed ($d(t)\leq u_{\R{max}}$), the controller builds and appropriately distributes the required pre-shock target inventory distribution $\mathbf{x}^\star$ obtained from \eqref{eq:optimization}; when a demand shock occurs ($d(t)>u_{\R{max}}$), the controller exploits the available inventory to sustain such demand for the prescribed survival horizon.

To account for the transportation delays in the chain, we define the \emph{inventory position} $p_i$ of each node as
\begin{equation} \label{eq:inventory_position}
    p_i(t) \coloneq \begin{cases}
        x_1(t), & i=1,\\
        x_i(t)+\int_{t-\tau}^{t} f_{i-1}(s)\,\R{d}s, & i=2,\ldots,N.
    \end{cases}
\end{equation}
Then, we construct the \emph{target inventory positions} $p_i^\star$ by considering the nominal operating condition in which the internal flows match the external demand:
\begin{equation}
    p_i^\star(t) \coloneq  \begin{cases}
        x_1^\star, & i=1, \\
        x_i^\star + \int_{t-\tau}^{t} d(s)\,\R{d}s, & i=2,\ldots,N,
    \end{cases}
\end{equation}
which, for a constant demand $d(t)=\bar{d}$, becomes $p_i^\star=x_i^\star + \bar{d}\tau$, for $i = 2,\ldots, N$.
Finally, we define $M_i$ and $M_i^\star$ as the total amount of commodity downstream of node $i$ and its corresponding target, respectively:
\begin{equation}\label{eq:cumulative_mass}
    M_i(t)\coloneq\sum_{j=i+1}^{N}p_j(t),
    \quad
    M_i^\star(t)\coloneq\sum_{j=i+1}^{N}p_j^\star(t),
\end{equation}
for $i=0,\ldots,N-1$. 
The case $i=0$ accounts for the total amount of commodity in the supply chain.
Defining the errors $e_i(t)\coloneq M_i^\star(t)-M_i(t)$, we propose the following two-mode control strategy:
\begin{equation}\label{eq:input_control}
u(t)= \begin{cases}
\mathrm{sat}_{[0,u_{\R{max}}]} \left(d(t)+\beta e_0(t)\right), & d(t)\leq u_{\R{max}}, \\
u_{\R{max}}, & d(t)>u_{\R{max}},
\end{cases}
\end{equation}
\begin{equation}\label{eq:flow_control}
f_i(t)= \begin{cases}
\mathrm{sat}_{[0,\beta x_i(t)]} \left(d(t)+\beta e_i(t)\right), & d(t)\leq u_{\R{max}}, \\
\beta x_i(t), & d(t)>u_{\R{max}},
\end{cases}
\end{equation}
for $i=1,\ldots,N-1$, where $\mathrm{sat}_{[a,b]}(z)\coloneq \min\{b,\max\{a,z\}\}$.

During periods of ordinary demand, the controller combines feedforward control with a proportional feedback action to regulate the distribution of commodity along the chain. 
During a demand shock, all flows are set to their maximum admissible rates.
Under this policy, the prescribed survival time is guaranteed if, at the onset of the shock, the inventories and flow histories satisfy the assumptions and conditions of Theorem~\ref{thm:inventory_conditions}.

The following result characterizes the closed-loop system during nominal operation.

\begin{proposition}\label{prop:closed_loop_equilibrium}
For constant demand $d(t)=\bar{\eta} \in (0,u_{\R{max}})$ and target inventories satisfying $\beta x_i^\star\geq u_{\R{max}}$, $i=1,\ldots,N-1$, system \eqref{eq:chain_dynamics} controlled with \eqref{eq:input_control}--\eqref{eq:flow_control} admits a unique equilibrium $\mathbf{x}=\mathbf{x}^\star$, with constant internal flow histories $\vec{\eta}_t = \bar{\eta}\vec{1}_{N-1}$. Moreover, this equilibrium is locally exponentially stable.
\end{proposition}

\begin{proof}
At $\mathbf{x}=\mathbf{x}^\star$, with constant flows equal to $\bar{\eta}$, all errors $e_i$ vanish. Since $\beta x_i^\star\geq u_{\R{max}}>\bar{\eta}$, the control laws yield $u=f_i=\bar{\eta}$, $\forall i$, establishing the existence of the equilibrium.

To prove uniqueness, consider that, for any equilibrium, mass conservation requires $u=f_i=\bar{\eta}$. Therefore, \eqref{eq:input_control} implies $e_0=0$ while \eqref{eq:flow_control} implies $e_i \ge 0$ for $i = 1, \ldots, N-1$. Specifically, suppose some $e_i>0$, and let $i$ be the smallest such index. Then $e_{i-1}=0$ and, since the flows are constant, $x_i=x_i^\star+e_i>x_i^\star$.
Consequently, both $\beta x_i$ and $\bar{\eta}+\beta e_i$ exceed $\bar{\eta}$, implying $f_i>\bar{\eta}$, a contradiction.
Thus all errors vanish, yielding $\mathbf{x}=\mathbf{x}^\star$.

Finally, it is straightforward to verify that the total downstream commodity quantities satisfy $\dot{M}_0(t) = u(t) - d(t)$ and $\dot M_i(t)=f_i(t)-d(t), \ i=1,\ldots,N-1$.
Since the saturations of \eqref{eq:input_control}--\eqref{eq:flow_control} are inactive at the equilibrium, they remain inactive for sufficiently small perturbations. In this regime, the control laws \eqref{eq:input_control}--\eqref{eq:flow_control} yield $\dot{e}_i(t)=-\beta e_i(t)$, and hence $e_i(t)=e_i(0)e^{-\beta t}$, for $i = 0, \ldots, N-1$.

Moreover, since $p_i^\star-p_i=e_{i-1}-e_i$, with $e_N=0$, the exponential decay of the errors implies exponential convergence of the inventory positions $p_i$ to their desired values $p_i^\star$. Similarly, \eqref{eq:flow_control} yields $f_i-\bar{\eta}=\beta e_i$, so the internal flows converge exponentially to $\bar{\eta}$ and consequently the inventories $x_i$ converge exponentially to their desired values $x_i^\star$.
\end{proof}

\section{Numerical validation}
\label{sec:numerical_validation}
\begin{figure*}[!t]
    \centering
    \captionsetup[subfigure]{skip=0pt, justification=raggedright, singlelinecheck=false, margin=0pt}

    % (a) Open-loop inventories
    \begin{subfigure}[t]{0.48\textwidth}
        \begin{minipage}[t]{0.05\linewidth}
            \vspace{10pt}
            \caption{}
            \label{fig:open_loop_inventory}
        \end{minipage}%
        \begin{minipage}[t]{0.95\linewidth}
            \vspace{0pt}
            \centering
            \includegraphics[width=0.9\linewidth]{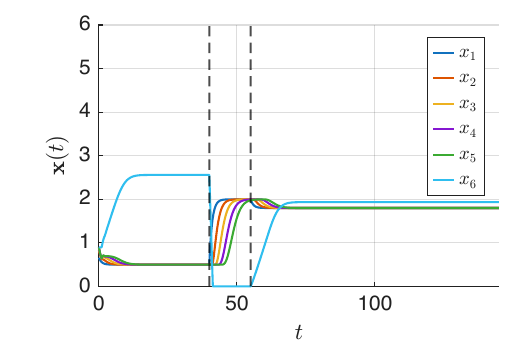}
        \end{minipage}
    \end{subfigure}%
    \hfill
    % (b) Closed-loop inventories
    \begin{subfigure}[t]{0.48\textwidth}
        \begin{minipage}[t]{0.05\linewidth}
            \vspace{10pt}
            \caption{}
            \label{fig:closed_loop_inventory}
        \end{minipage}%
        \begin{minipage}[t]{0.95\linewidth}
            \vspace{0pt}
            \centering
            \includegraphics[width=0.9\linewidth]{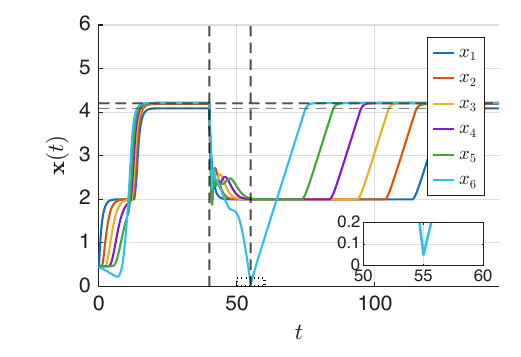}
        \end{minipage}
    \end{subfigure}

    \par\nointerlineskip

    % (c) Open-loop input and demand
    \begin{subfigure}[t]{0.48\textwidth}
        \begin{minipage}[t]{0.05\linewidth}
            \vspace{10pt}
            \caption{}
            \label{fig:open_loop_demand_and_input}
        \end{minipage}%
        \begin{minipage}[t]{0.95\linewidth}
            \vspace{0pt}
            \centering
            \includegraphics[width=0.9\linewidth]{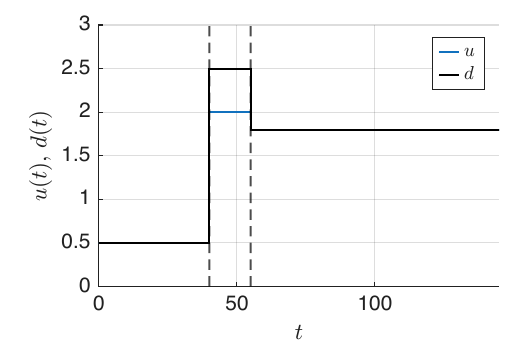}
        \end{minipage}
    \end{subfigure}%
    \hfill
    % (d) Closed-loop input and demand
    \begin{subfigure}[t]{0.48\textwidth}
        \begin{minipage}[t]{0.05\linewidth}
            \vspace{10pt}
            \caption{}
            \label{fig:closed_loop_demand_and_input}
        \end{minipage}%
        \begin{minipage}[t]{0.95\linewidth}
            \vspace{0pt}
            \centering
            \includegraphics[width=0.9\linewidth]{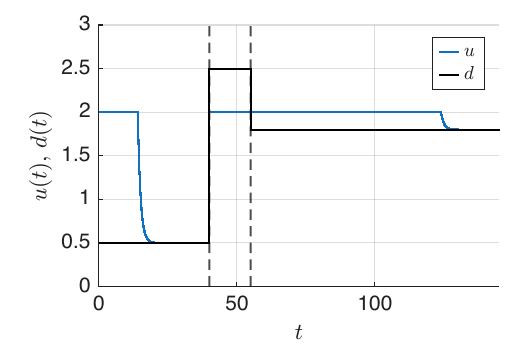}
        \end{minipage}
    \end{subfigure}

    \caption{Time evolution of the inventories (a), (b) and external input and demand (c), (d). Left column: baseline policy with $u(t)=\min\{d(t),u_{\max}\}$ and $f_i(t)=\beta x_i(t)$. Right column: proposed control strategy (25)–(26). Vertical dashed lines indicate the demand shock onset ($t=40$) and end ($t=55$), while horizontal dashed lines in (b) indicate the target inventory levels $\vec{x}^\star$.}
    \label{fig:comparison}
\end{figure*}

In this section, we numerically investigate the response of the supply chain \eqref{eq:chain_dynamics}--\eqref{eq:outflow_rate} to a constant demand shock and evaluate the effectiveness of the proposed control strategy. We consider a network of $N=6$ nodes, with transportation delay $\tau=1$, maximum release rate $\beta=1$, maximum external input $u_{\R{max}}=2$, minimum demand $d_{\R{min}} = 0.5$, and maximum demand $d_{\R{max}} = 2.5$. 
The inventory targets $\vec{x}^\star$ are computed solving \eqref{eq:optimization} with holding costs $h_i=1$, $i = 1,\ldots,N$, number of samples $L_i=200$, $i = 1,\ldots,N-1$, margin $\varepsilon=0.05$, and lower bound on the exit time $\bar{T}=15$.
The shock occurs at $t_0=40$ and lasts $T_{\mathrm s}=15$, as shown in Figs.~\ref{fig:open_loop_demand_and_input}--\ref{fig:closed_loop_demand_and_input}.

First, as a baseline, we consider the supply chain operating under the linear push laws \eqref{eq:chain_dynamics_push} with external input $u(t)=\min\{d(t),u_{\R{max}}\}$. This policy matches the external input to the demand whenever feasible without regulating inventory levels. As shown in Fig.~\ref{fig:open_loop_inventory}, following the shock, the input saturates and the available inventories cannot sustain the increased demand. Consequently, the last inventory reaches zero before the shock ends, causing the system to leave $\mathcal{X}_{\mathrm{op}}$.

Next, we apply the control strategy \eqref{eq:input_control}--\eqref{eq:flow_control} to the same supply chain. As shown in Fig.~\ref{fig:closed_loop_inventory}, the inventories converge to their prescribed targets before the shock, the last inventory remains strictly positive throughout the shock, and all inventories progressively return to their target levels afterward. Unlike the baseline policy, the proposed strategy thus ensures demand satisfaction for the prescribed time horizon.

\section{Conclusions}
\label{sec:conclusions}
We investigated finite-horizon demand satisfaction in delayed serial supply chains subject to demand shocks exceeding the maximum input capacity. 
We derived an analytical characterization of the inventory trajectories under a specified flow policy and established necessary and sufficient conditions for guaranteeing a prescribed survival time. 
These conditions were used to determine suitable inventory reserves through convex optimization and to design a two-mode controller for nominal and emergency operation.
Numerical simulations illustrated the effectiveness of the proposed approach in improving the ability of the supply chain to withstand demand shocks.

Future work will focus on extending these results to different flow control policies such as pull-like flows and to different network topologies.

\bibliographystyle{IEEEtran}
\bibliography{references}

\end{document}